\documentclass[pra,aps,twocolumn,showpacs]{revtex4}
\usepackage{amsbsy}
\usepackage{bm}
\usepackage{graphicx}

\newcommand{\nc}{\newcommand}
\nc{\eq}[1]{\mbox{Eq.~(\ref{#1})}}
\nc{\ba}{\begin{array}}
\nc{\ea}{\end{array}}
\nc{\bea}{\begin{eqnarray}}
\nc{\eea}{\end{eqnarray}}
\nc{\fig}[1]{\mbox{Fig.~\ref{#1}}}

\begin{document}

\title{Thermal spin flips in atom chips}
\author{P.K.~Rekdal}
\author{S.~Scheel}
\email{s.scheel@imperial.ac.uk}
\author{P.L.~Knight}
\author{E.A.~Hinds}
\affiliation{Quantum Optics and Laser Science, Blackett Laboratory,
Imperial College London, Prince Consort Road, London SW7 2BW, United Kingdom}

\date{\today}

\begin{abstract}
We derive a general expression for the spin-flip rate of an atom
trapped near an arbitrary dielectric body and we apply this theory
to the case of a $2$-layer cylindrical metal wire. The spin flip
lifetimes we calculate are compared with those expected for an
atom near a metallic slab and with those measured by Jones {\it et
al.} above a 2-layer wire [M.P.A. Jones, C.J. Vale, D. Sahagun,
B.V. Hall, and E.A. Hinds, Phys. Rev. Lett. {\bf 91}, 080401
(2003)]. We investigate how the lifetime depends on the skin depth
of the material and on the scaling of the dimensions. This leads
us to some conclusions about the design of integrated circuits for
manipulating ultra-cold atoms (atom chips).
\end{abstract}


\maketitle

\section{Introduction}

Microscopic traps provide a powerful tool for the control and
manipulation of  Bose-Einstein condensates over micrometer
distances. Microstructured surfaces, known as atom chips, are
particularly interesting for this purpose since they can be
tailored to provide a variety of trapping geometries
\cite{weinstein_95} and promise well-controlled quantum state
manipulations of neutral atoms in integrated and scalable
microtrap arrays. Ultimately there is the possibility of
controlling the quantum coherences within arrays of individual
atoms for use in quantum information processing \cite
{calarco_00}. This technology is attractive because it appears
robust and scalable and because trapped neutral atoms can have
long coherence times.

However, atoms in these traps are held close to the
micro-structured material surfaces, which are typically at room
temperature. Thermal fluctuations give rise to Johnson noise
currents in the material \cite{johnson_28}. Such currents are
normally observed as a noise voltage across a resistor, but they
also cause the electromagnetic field near a conducting solid to
fluctuate with a broad noise spectrum. For atoms trapped close to
the surface of a conductor these fluctuating fields can be strong
enough to drive rf magnetic dipole transitions that flip the
atomic spin. If the atom is in a magnetic trap where only
low-field-seeking Zeeman sublevels are confined, the spin flips
lead to atom loss. This is known experimentally
\cite{hinds_03,harber_03} as well as theoretically
\cite{henkel_99}. The loss rate increases strongly as the atoms
approach the metallic surface of an atom chip. For a given desired
lifetime, this restricts how close the trapped atoms can be
brought to the surface, which in turn determines the period of the
smallest trapping structures that can be imposed on the atom by
the chip.

The paper is organized as follows: In Section \ref{Bas_eq_Q} we
introduce the basic equations and discuss the quantization of an
electromagnetic field in the presence of a dispersing and
absorbing dielectric body. Then, in Section \ref{gamma_sec}, we
derive a general expression for the spontaneous and thermal
spin-flip rates of an atom due to the coupling of its magnetic
moment to the magnetic field. This derivation is based on the
Zeeman Hamiltonian of the system and the corresponding Heisenberg
equations of motion. We show that the spin-flip rate is determined
by the dyadic Green tensor of the classical, phenomenological
Maxwell equations. In Section \ref{G_Sec} we present the
scattering Green tensor for a $2$-layer cylindrical body
surrounded by an unbounded homogeneous medium, with details being
given in Appendix \ref{R_appendix}. Then, in Section \ref{final
formula}, we use this Green tensor to obtain an explicit
analytical expression for the total spin-flip rate of an atom
above a $2$-layer wire. Some numerical results are presented and
discussed in Section \ref{Num_Sec}.  The numerical results are
compared with the corresponding results for a slab and with the
experimental measurements presented by Jones et al. in
Ref.~\cite{hinds_03}. Our conclusions are given in Section
\ref{conclusions}.


\section{Basic equations and quantization}
\label{Bas_eq_Q}

In classical electrodynamics, dielectric matter is commonly described
in terms of a phenomenologically introduced dielectric susceptibility.
Let us consider a classical electromagnetic field, described by the
phenomenological Maxwell's equations, without external sources.
We restrict our attention to isotropic but arbitrarily
inhomogeneous non-magnetic media, and assume that the
polarization responds linearly and locally to the electric field.
A linear response formalism similar to that presented below can also
be found in Refs. \cite{agarwal_75,dung_00}.

The most general relation between the matter polarization and the
electric field consistent with causality and the
fluctuation-dissipation theorem is \cite{landau_60}
\bea
\label{P_rt} {\bf P}({\bf r},t) = \varepsilon_0
\int\limits_0^{\infty} d \tau \, \chi({\bf r},\tau)\, {\bf E}({\bf
r},t-\tau) +{\bf P}_N({\bf r},t)\,,
\eea
where $\chi({\bf r},t)$
is the linear susceptibility. The inclusion of the noise
polarization ${\bf P}_N({\bf r},t)$ is necessary to fulfil the
fluctuation-dissipation theorem. It is this fluctuating part of
the polarization that is unavoidably connected with the loss in
the medium. Converting the displacement field ${\bf D}({\bf
r},t)=\varepsilon_0 {\bf E}({\bf r},t) + {\bf P}({\bf r},t)$ into
Fourier space using \eq{P_rt}, we obtain \bea
   {\bf D}({\bf r},\omega) = \varepsilon_0  \varepsilon({\bf r},\omega)
   {\bf E}({\bf r},\omega)  + {\bf P}_N({\bf r},\omega) \,,
\eea where $\varepsilon({\bf r},\omega)$ is the complex
permittivity and $\varepsilon({\bf r},\omega)-1$ is the temporal
Fourier transform of $\chi({\bf r},t)$. The real part of the
permittivity ($\varepsilon_R$, responsible for dispersion) and
the imaginary part ($\varepsilon_I$, responsible for absorption)
are related to each other by the Kramers-Kronig relation.

Using Maxwell's equations in Fourier space, we find that
$\textbf{E}({\bf r},\omega)$ satisfies  the Helmholtz equation
\begin{equation}
\nabla \times \nabla \times   {\bf E}({\bf r},\omega)
-\frac{\omega^2}{c^2}
\varepsilon({\bf r},\omega) {\bf E}({\bf r},\omega )
= \omega^2\mu_0 {\bf P}_N({\bf r},\omega) \, ,
\end{equation}
with the solution \bea {\bf E}({\bf r},\omega)  = \omega^2  \mu_0
\int d^3 {\bf r}' \, \bm{G}({\bf r},{\bf r}',\omega ) \cdot {\bf
P}_N({\bf r}',\omega) \,, \eea where the Green tensor $\bm{G}({\bf
r},{\bf r}',\omega)$ is a second-rank tensor determined by the
partial differential equation
\begin{equation}   \label{G_Helm}
\nabla\times\nabla\times \bm{G}({\bf r},{\bf r}',\omega)
- \frac{\omega^2}{c^2}
\varepsilon({\bf r},\omega) \bm{G}({\bf r},{\bf r}',\omega)
= \delta( {\bf r} - {\bf r}' ) \bm{U} \, ,
\end{equation}
where $\bm{U}$ is the unit dyad.
Together with the boundary condition at infinity, this equation
has a unique solution. In accordance with Maxwell's equations the
corresponding solution for the magnetic field in Fourier space is
${\bf B}({\bf r},\omega)=(i\omega)^{-1} \nabla\times {\bf E}({\bf
r},\omega)$ .

As we have seen, the noise polarization ${\bf P}_N({\bf r},t)$
plays a fundamental r\^{o}le in determining the electric field.
The form of ${\bf P}_N({\bf r},t)$ follows from the
fluctuation-dissipation theorem, which states that the
fluctuations of the macroscopic polarization are given by the
imaginary part of the response function [here $\varepsilon_I({\bf
r},\omega)$]. If we pull out a factor and define the dynamical
variables ${\bf f}({\bf r},\omega)$ as the fundamental
$\delta$-correlated random process, we find that we can write the
noise polarization as \cite{dung_00}
 \bea \label{P_N} {\bf
P}_N({\bf r},\omega) = i \; \sqrt{\frac{\hbar\varepsilon_0}{\pi}
\, \varepsilon_I({\bf r},\omega)} \; {\bf f}({\bf r},\omega) \,.
\eea Upon quantization, we replace the classical fields ${\bf
f}({\bf r},\omega)$ by the operator-valued bosonic fields
$\hat{\bf f}({\bf r},\omega)$ which we associate with the
elementary excitations of the system composed of the
electromagnetic field and the absorbing dielectric matter. They
satisfy the equal-time commutation relations $[ \hat{f}_i({\bf
r},\omega),\hat{f}^\dagger_j({\bf r}',\omega')]$ $= \delta_{ij}
\delta({\bf r}-{\bf r}')\delta(\omega-\omega')$.

The magnetic-field operator in the Schr\"{o}dinger picture can now be
obtained as
\bea
\hat{\bf B}({\bf r}) =  \hat{\bf B}^{(+)}({\bf r})
+ \hat{\bf B}^{(-)}({\bf r}) \,,\quad
 \hat{\bf B}^{(-)}({\bf r}) =[\hat{\bf B}^{(+)}({\bf r})]^\dagger
\eea
where
\bea
  \hat{\bf B}^{(+)}({\bf r}) = \int\limits_0^{\infty} d \omega \,
  \hat{\bf B}({\bf r},\omega) \; ,
\eea
is its positive-frequency part. In this way, the
electromagnetic field is expressed in terms of the classical Green
tensor satisfying  the Helmholtz equation (\ref{G_Helm}) and the
continuum of the fundamental bosonic field variables $\hat{\bf
f}({\bf r},\omega)$. All the information about the dielectric
matter is contained, via the permittivity $\varepsilon({\bf
r},\omega)$,  in the Green tensor of the classical problem.

We close this Section by mentioning two important properties of the
Green tensor.
It can be shown that the (Onsager) reciprocity relation
$\bm{G}({\bf r},{\bf r}',\omega)=\bm{G}^T({\bf r}',{\bf r},\omega)$
holds \cite{onsager_31}. Additionally, another useful property is the
integral relation
\bea  \label{G_rel}
\int d^3 {\bf r}' \; \frac{\omega^2}{c^2} \,
\varepsilon_I({\bf r}',\omega) \,
G_{kl}({\bf r},{\bf r}',\omega )
G_{nl}^\ast({\bf r}_A,{\bf r}',\omega )
\\ \nonumber
= \mbox{Im} \,G_{kn}({\bf r},{\bf r}_A,\omega) \, ,&&
\eea
which we will use later in this paper. Both relations essentially
follow from linear response theory, with \eq{G_rel} being equivalent
to the fluctuation-dissipation theorem \cite{Eckhardt}.
It should be noted that we assume the dielectric permittivity to
possess at least an infinitesimal imaginary part everywhere to avoid
surface contributions in Eq.~(\ref{G_rel}).


\section{Derivation of the spontaneous and thermal spin flip rates}
\label{gamma_sec}

The Hamiltonian of the combined system of electromagnetic field and
absorbing matter, from which the (quantized) phenomenological
Maxwell's equations can be derived, can be written in terms of the
basic field operators $\hat{\bf f}({\bf r},\omega)$ in the diagonal
form
\begin{equation}
\label{H_free}
\hat{H} =  \int d^3 {\bf r}  \int\limits_{0}^{\infty} d \omega \, \hbar
\omega \; \hat{\bf f}^{\dagger}({\bf r},\omega) \cdot
\hat{\bf f}({\bf r},\omega)
+\sum_{\alpha = i,f} \hbar \omega_{\alpha}  \hat{\xi}_{\alpha}  \,,
\end{equation}
which leads  in the Heisenberg picture to the (quasi-free) time
evolution $\hat{\bf f}({\bf r},\omega)$ $\rightarrow$ $\hat{\bf
f}({\bf r},\omega)  e^{- i \omega t}$. Here we have also included
an atom through the operators $\hat{\xi}_{\alpha}
\equiv | \alpha \rangle \langle \alpha |$ and the energy $\hbar
\omega_{\alpha}$  of the atomic state $| \alpha \rangle$ ($\alpha
= i,f$).

The interaction of the atom at position ${\bf r}_A$ with a
magnetic field $\hat{{\bf B}}({\bf r})$ is described by the Zeeman
Hamiltonian $\hat{H}_{\text{Z}} = - \hat{\bm \mu} \cdot \hat{{\bf
B}}({\bf r}_A)$, where $\hat{\bm \mu} = {\bm \mu} | i \rangle
\langle f | + \text{h.c.}$ is the magnetic moment operator
associated with the transition  $|i \rangle \rightarrow |f
\rangle$. The magnetic moment vector is \bea {\bm \mu} =  \langle
i | \; \mu_B \bigg ( g_S  \hat{\bf S} + g_L  \hat{\bf L} - g_I
\frac{m_e}{m_p} \hat{\bf I} \bigg ) \,  | f \rangle \; , \eea
where $\mu_B$ is the Bohr magneton, $\hat{\bf S}$ is the
electronic spin operator, $\hat{\bf L}$ is the orbital angular
momentum operator, $\hat{\bf I}$ is the nuclear spin operator and
$g_S\approx 2$, $g_L$ and $g_I$ are the corresponding $g$-factors.
We restrict our attention to $L=0$, which corresponds to the
ground state of an alkali atom, and we neglect the small nuclear
magnetic moment in comparison with the Bohr magneton. In the
rotating-wave approximation, we can then write the Zeeman
Hamiltonian as
\begin{equation} \label{H_RWA}
\hat{H}_Z \approx -\mu_B g_S \left[ \; \langle
f|\hat{S}_q|i\rangle \, \hat{\xi}^{(+)} \hat{B}_q^{(+)}({\bf r}_A)
+\mbox{h.c.} \; \right] \,,
\end{equation}
where the atomic raising (lowering) operator
$\hat{\xi}^{(+)} \equiv |i \rangle \langle f |$ [$\hat{\xi}^{(+)}
= ( \hat{\xi}^{(-)} )^{\dagger}$] satisfies the commutation
relation $[ \, \hat{\xi}_z,\hat{\xi}^{(\pm)}]=\pm
\hat{\xi}^{(\pm)}$, with $\hat{\xi}_z \equiv$
$\frac{1}{2}(|i\rangle\langle i |-|f \rangle\langle f|)$. Repeated
indices $q$ indicate a sum over spatial vector components.

Using the Hamiltonian (\ref{H_RWA}), the Heisenberg equation of
motion for the atomic quantity $\hat{\xi}_z(t)$ is given by
\bea
\label{mu_DL} \dot{\hat{\xi}}_z(t)  = - \frac{\mu_B g_S}{i\hbar}
\, \langle f|\hat{S}_q|i\rangle \; \hat{\xi}^{(+)} \,
\hat{B}_q^{(+)}({\bf r}_A) +\mbox{h.c.} \, .
\eea
Furthermore, the Heisenberg equation of motion for the bosonic field
operator is 
\bea \label{f_DL}
\dot{ \hat{f}_i}({\bf r},\omega,t)&=& - i\omega
\hat{f}_i({\bf r},\omega, t)  \\ \nonumber &&   \hspace*{-18ex}
+\frac{i\mu_B g_S}{\sqrt{\hbar\pi\varepsilon_0}} \,
\langle i|\hat{S}_q|f \rangle  \, \hat{\xi}^{(-)} \epsilon_{qpj}
\partial_p \frac{\omega}{c^2} \sqrt{ \varepsilon_I({\bf r},\omega)
} \, G_{ji}^\ast({\bf r}_A,{\bf r},\omega ) \; ,
\eea
where
$\epsilon_{qpj}$ is the Levi-Civita symbol and $\partial_j \equiv
\partial/\partial x_j$. This equation can now be formally
integrated to yield
\bea \label{f_markov}
\lefteqn{ \hat{f}_i({\bf r},\omega,t)
=\hat{f}_{i,\text{free}}({\bf r},\omega,t) +
\int\limits_{t'}^{t} d \tau \, e^{-i\omega(t-\tau)}
\hat{\xi}^{(-)}(\tau) }
\\ && \nonumber
\times \frac{i\mu_B g_S}{\sqrt{\hbar\pi\varepsilon_0}} \,
\langle i|\hat{S}_q|f \rangle \,
\epsilon_{qpj} \, \partial_p \frac{\omega}{c^2}
\sqrt{ \varepsilon_I({\bf r},\omega) } \,
G_{ji}^\ast({\bf r}_A,{\bf r},\omega ) \, ,
\eea
where $\hat{f}_{i,\text{free}}({\bf r},\omega,t)$ denotes the freely
evolving basic-field operators.
The lowering operator $\hat{\xi}^{(-)}(\tau)$ in \eq{f_markov} can
be found by solving its Heisenberg equation of motion. In the
Markov approximation, this solution can be reduced to its slowly
varying part $\hat{\xi}^{(-)}(t) \, e^{i \omega_{if} (t - \tau)}$
in \eq{f_markov} so that the time integral can be approximated by
\bea
&& \hat{f}_i({\bf r},\omega,t)
=\hat{f}_{i,\text{free}}({\bf r},\omega,t)
+\frac{i\mu_Bg_S}{\sqrt{\hbar\pi\varepsilon_0}} \,
\langle i|\hat{S}_q|f\rangle \; \hat{\xi}^{(-)}(t) \nonumber \\ &&
\times \epsilon_{qpj} \, \partial_p \frac{\omega}{c^2} \sqrt{
\varepsilon_I({\bf r},\omega) } \, G_{ji}^\ast({\bf r}_A,{\bf
r},\omega) \, \zeta(\omega_{if}-\omega) \, ,
\eea
where
$\zeta(x)=\pi\delta(x)+i \, {\cal P}x^{-1}$ ($\cal P$ denotes the
principal value) and $\omega_{if} \equiv \omega_i - \omega_f$ is
the transition frequency corresponding to the flip $|i \rangle
\rightarrow |f \rangle$ in the atom's internal state. Substituting
this formal solution into the expression for the magnetic field,
we obtain
\bea \label{B_sol}
\lefteqn{ \hat{B}_q^{(+)}({\bf r}_A,\omega,t) =
\hat{B}_{q,\text{free}}^{(+)}({\bf r}_A,\omega,t)
} \nonumber \\ && + \frac{i\mu_B g_S \mu_0}{\pi} \,
\langle i|\hat{S}_p|f\rangle \epsilon_{qjk} \epsilon_{pmn}
\partial_j \partial_m \, \hat{\xi}^{(-)}(t) \, \zeta(\omega_{if}-\omega)
\nonumber \\ && \times \int d^3{\bf r}\, \frac{\omega^2}{c^2} \,
\varepsilon_I({\bf r},\omega) \, G_{kl}({\bf r}_A,{\bf r}, \omega)
G_{nl}^\ast({\bf r}_A,{\bf r},\omega) \,. \eea
The spatial integral can be evaluated using the integral relation
\eq{G_rel} yielding $\text{Im}\,G_{kn}({\bf r}_A,{\bf r}_A,\omega)$.
Therefore, \eq{B_sol} becomes
\bea \lefteqn{ \hat{B}_q^{(+)}({\bf r}_A,\omega,t) =
\hat{B}_{q,\text{free}}^{(+)}({\bf r}_A,\omega,t)
} \nonumber \\ && + \frac{i\mu_B g_S \mu_0}{\pi} \,
\langle i|\hat{S}_k|f \rangle \, \hat{\xi}^{(-)}(t) \;
\zeta(\omega_{if}-\omega) \nonumber \\ && \times \; \text{Im}[ \;
\nabla\times\nabla\times \bm{G}({\bf r}_A,{\bf r}_A,\omega) \;
]_{qk} \; . \eea
Performing the $\omega$-integration and inserting
into \eq{mu_DL}, we obtain
\bea \lefteqn{ \dot{\hat{\xi}}_z(t)  =
-   ( \, \Gamma^B + i \delta \omega \, ) \, [ \; \frac{1}{2} +
\hat{\xi}_z(t) \; ] }
 \nonumber \\ &&
+ \bigg [ \, \frac{i\mu_B g_S}{\hbar} \, \langle
f|\hat{S}_q|i\rangle \, \hat{\xi}^{(+)} \,
\hat{B}_{q,\text{free}}^{(+)}({\bf r}_A) +\mbox{h.c.} \, \bigg ]
\, , \eea
where the spontaneous spin-flip rate $\Gamma^B \equiv
\Gamma^B({\bf r}_A, \tilde{\omega}_{if})$ arises from the $\delta$
function (the real part of the $\zeta$ function) and is given by
\bea \label{gamma_B_generall} \lefteqn{ \Gamma^B = \, \mu_0 \,
\frac{2 \, (\mu_B g_S)^2}{\hbar} \; \langle f|\hat{S}_q|i \rangle
\langle i|\hat{S}_p|f \rangle } \nonumber \\ && \; \times \;
\mbox{Im}  \, [  \; \nabla \times \nabla \times
  \bm{G}({\bf r}_A  ,  {\bf r}_A  ,  \tilde{\omega}_{if} ) \; ]_{qp}  \; ,
\eea
and where the term $\delta \omega$ arises from the principal-value
integral (the imaginary part of the $\zeta$ function) and is
identified as the radiative frequency shift. Furthermore, the
shifted frequency is given by
$\tilde{\omega}_{if}=\omega_{if}+\delta\omega$. In what follows,
the transition frequency is always taken to be the shifted
frequency $\tilde{\omega}_{if}$ that one measures in an experiment
and not the bare frequency $\omega_{if} \equiv \omega$. For
simplicity, we omit the tilde in all subsequent formulas. Note
that the same result for $\Gamma^B$ is obtained when using an
appropriately derived master equation as done in
Ref.~\cite{henkel_99}.

We assume that the dielectric body is in thermal equilibrium with
its surroundings. The magnetic field  is then in a thermal state
with a temperature $T$, equal to the temperature of the dielectric
body. The total flip rate for the atom is therefore given by
$\Gamma^B_{\text{total}} =  \Gamma^B  ( \overline{n}_{\text{th}} +
1 )$, where the mean thermal occupation number is \bea
\label{n_th} \overline{n}_{\text{th}} = \frac{1}{e^{\hbar
\omega_{if}/k_{\text{B}} T}-1} \, , \eea and $k_{\text{B}}$ is
Boltzmann's constant. At zero temperature, i.e.
$\overline{n}_{\text{th}} = 0$, the relaxation dynamics is
entirely due to the spontaneous flip rate $\Gamma^B$. For large
$T$ on the other hand,  $\overline{n}_{\text{th}} \approx$
$\frac{k_{\text{B}} T}{\hbar \omega_{if}} \gg 1$ and the spin flip
rate is predominantly induced by thermal fluctuations.

In the experiment of Ref.~\cite{hinds_03} $^{87}$Rb atoms are
initially pumped into the trapped state $|F,m\rangle=|2,2\rangle$.
Thermal fluctuations of the magnetic field then cause the atoms to
evolve into hyperfine sublevels with lower $m_F$. Upon making a
transition to the $m_F=1$ state, the atoms are more weakly trapped
and are largely lost from the region of observation, causing the
measured atom number to decay with a rate $\Gamma^B_{21}$. Here we
are introducing the notation $\Gamma^B_{m_i m_f}$ for the total
spin-flip rate associated with the transition $|2,m_i\rangle
\rightarrow |2,m_f\rangle$.

\section{The dyadic Green tensor}
\label{G_Sec}

\begin{figure}[ht]
\centerline{\includegraphics[width=4.5cm]{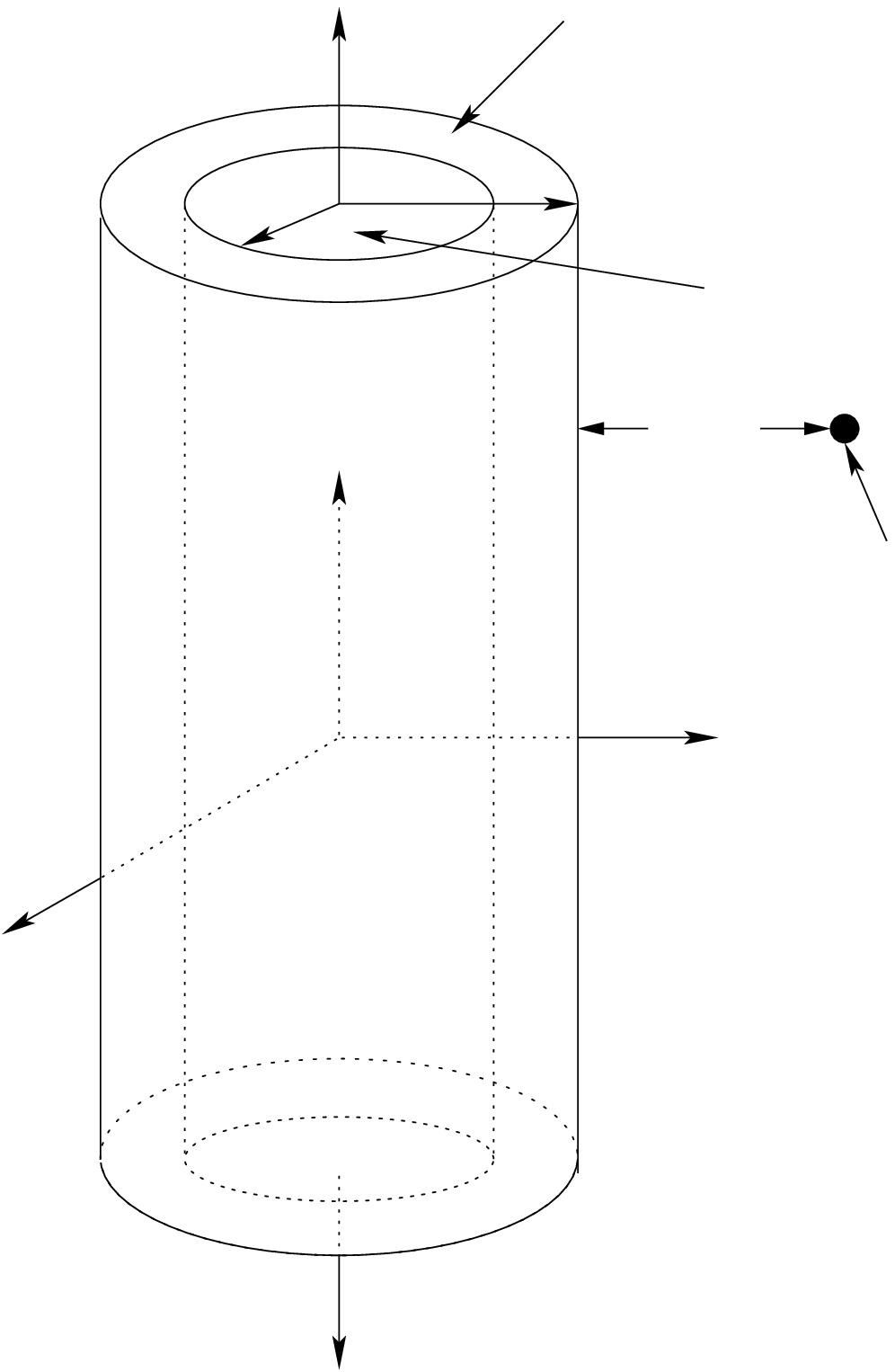}}
\unitlength=1mm
\begin{picture}(0,0)
\put(-4,64){\tiny $a_2$} \put(-10,63){\tiny $a_1$}
\put(14.8,57.6){layer $1$} \put(8,72){layer $2$}
\put(-35,70){layer $3$} \put(-36,67){(vacuum)} \put(-25,25){$x$}
\put(16,35){$y$} \put(-6,51){$z$} \put(13,51){$r$}
\put(18,43){\small atom}
\end{picture}
\caption{ The geometry we are considering is a $2$-layer cylinder
surrounded by an unbounded homogeneous medium. The outer region is
labelled layer $3$ (vacuum), the coating is layer $2$ and the
cylinder core is layer $1$.  The distance from the surface of the
outermost layer to the atom is $r$.} \label{cc_wire_FIG}
\end{figure}

The geometry we are considering in this paper is a $2$-layer cylinder
surrounded by an unbounded homogeneous medium (see
\fig{cc_wire_FIG}). This corresponds to the experimental geometry in
Ref. \cite{hinds_03}.
Because the Helmholtz equation is linear, the associated Green
tensor can be written as a sum, \bea \label{G_tot}
  \bm{G}({\bf r},{\bf r}',\omega)=
  \bm{G}^0({\bf r},{\bf r}',\omega)
+ \bm{G}^{\text{wire}}({\bf r},{\bf r}',\omega) \,, \eea where
$\bm{G}^0({\bf r},{\bf r}',\omega)$ represents the contribution
from the vacuum and $\bm{G}^{\text{wire}}({\bf r},{\bf
r}',\omega)$ describes the part due to the wire. When the atom is
located in layer 3, the scattering contribution  is \cite{chew_90}
\begin{equation}
\label{G_33} {\bm G }^{\text{wire}}({\bf r},{\bf r}', \omega)
= \frac{i}{8 \pi} \int_{-\infty}^{\infty} d h
 \sum_{n=0}^{\infty}  \frac{2-\delta_{0n}}{\eta_3^2} \;
 \bm{{\cal R}}_{n}(h) \, ,
\end{equation}
where
\bea \label{cal_R_n}
\lefteqn{
\bm{{\cal R}}_n(h) =}
\nonumber \\ &&
R_n^{11}(h)
 \bigg [ {\bf N}_{^e  n}^{(1)}(h) {\bf N}_{^e n}^{'(1)}(-h)
+  {\bf N}_{_o  n}^{(1)}(h) {\bf N}_{_o  n}^{'(1)}(-h)
\bigg ]  \nonumber \\ &+&
R^{12}_n(h)  \left( - \frac{\omega \varepsilon_3}{k_3}  \right)
\\ && \times \bigg [ \,
{\bf N}_{^e  n}^{(1)}(h) {\bf M}_{_o  n}^{'(1)}(-h)
-{\bf N}_{_o  n}^{(1)}(h) {\bf M}_{^e  n}^{'(1)}(-h)
\nonumber \\ &&
+{\bf M}_{^e  n}^{(1)}(h) {\bf N}_{_o  n}^{'(1)}(-h)
-{\bf M}_{_o  n}^{(1)}(h) {\bf N}_{^e  n}^{'(1)}(-h) \,
\bigg ] \nonumber
\\ &+& R^{22}_n(h) \nonumber
\bigg [ {\bf M}_{^e  n}^{(1)}(h) {\bf M}_{^e  n}^{'(1)}(-h) +{\bf
M}_{_o  n}^{(1)}(h) {\bf M}_{_o  n}^{'(1)}(-h) \bigg ] \,. \eea
For simplicity, we have omitted the tensor product symbol
$\otimes$ between the {\it e}ven and {\it o}dd cylindrical vector
functions defined by ${\bf M}_{^e _o n}(h) = \nabla \times [
\psi_{^e _o n}(h) {\bf z} ]$ and ${\bf N}_{^e _o
n}(h)=\nabla\times\nabla\times [\psi_{^e _o n}(h){\bf z}]/k_3$.
The scalar eigenfunctions $\psi_{^e _o n}(h)$ satisfy the
homogeneous scalar wave equation \cite{chew_90}. It follows from
these definitions that \bea \label{M_def} \nabla \times {\bf
M}_{^e _o n}(h) &=& k_3 {\bf N}_{^e _o  n}(h)  \; ,
\\
\label{N_def} \nabla \times {\bf N}_{^e _o  n}(h) &=& k_3  {\bf
M}_{^e _o  n}(h)  \, .
\eea
Explicitly,
 \bea\label{N_vec}
\lefteqn{ {\bf N}_{^e _o  n}(h)  = \frac{1}{k_3}
\bigg [  i h \frac{d Z_n(\eta_3 \rho)}{d \rho} \ba{rr} \cos  \\
\sin \ea (n \phi) {\bf e}_{\rho}  }  \\ && \hspace*{-3ex} \mp  ih
\frac{n}{\rho}  Z_n(\eta_3 \rho) \ba{rr} \sin  \\ \cos \ea (n
\phi) {\bf e}_{\phi} + \eta_3^2   Z_n(\eta_3 \rho) \ba{rr} \cos
\\ \sin \ea (n \phi) {\bf e}_z  \bigg ]
 e^{i h z} \,,\nonumber
\eea
\bea
\label{M_vec}
\lefteqn{
{\bf M}_{^e _o  n}(h)
=
\bigg [  \mp  \frac{n}{\rho}  Z_n(\eta_3 \rho)
\ba{rr} \sin  \\ \cos \ea (n \phi)  {\bf e}_{\rho} } \nonumber \\ &&
\; - \, \frac{d Z_n(\eta_3 \rho)}{d \rho}
\ba{rr} \cos  \\ \sin \ea (n \phi)  {\bf e}_{\phi} \bigg ]
 e^{i h z} \,.
\eea
The primes in \eq{cal_R_n} indicate the spherical coordinates
$(\rho',\phi',z')$.
The superscript $(1)$ indicates that $Z_n$ should be replaced by the
Hankel function of first kind $H_n^{(1)}$. Otherwise, $Z_n$ is the
Bessel function  of first kind $J_n$.
The propagation constant in the $\rho$ direction is $\eta_p^2 = k_p^2-h^2$,
where $k_p$ is the wave number of the $p$th layer.
The permittivity of the $p$th layer is denoted by $\varepsilon_p$.
The scattering reflection coefficients $R^{kl}_n(h)$ are given in
Appendix \ref{R_appendix} ($k,l=1,2$).

The double curl of the Green tensor in \eq{G_33} can be written
\bea \label{curl_curl_G}
&& \nabla \times \nabla' \times
{\bf G }^{\text{wire}}({\bf r},{\bf r}', \omega) =
\\ \nonumber
&&
\frac{i}{8 \pi} \sum_{n=0}^{\infty} ( 2-\delta_{0n} )
\label{2_curl}
\left[
\ba{rrr}
(I_n)_{xx} & (I_n)_{xy} & (I_n)_{xz}  \\
(I_n)_{yx} & (I_n)_{yy} & (I_n)_{yz}  \\
(I_n)_{zx} & (I_n)_{zy} & (I_n)_{zz} \ea \right]  , \eea where
\bea  \label{2_curl_G_sum} {\bf I}_{ n}  \equiv {\bf I}_{n} ({\bf
r},{\bf r}', \omega) =\int_{-\infty}^{\infty}  d h   \;
\frac{1}{\eta_3^2} \; \nabla \times \nabla' \times  \bm{{\cal
R}}_{n}(h)  \; . \eea Note that the curls are computed by
replacing ${\bf N}_{^e _o  n}(h)$ by ${\bf M}_{^e _o  n}(h)$ and
vice versa, according to Eqs. (\ref{M_def}) and (\ref{N_def}).
Also note that the integration variable $h$ is the wave number in
the $z$-direction (see \fig{cc_wire_FIG}). From the symmetry of
the integrand, it is easy to show that
$(I_n^{\text{lim}})_{xz}=(I_n^{\text{lim}})_{zx}=(I_n^{\text{lim}})_{yz}=(I_n^{\text{lim}})_{zy}=0$,
where $(I_n^{\text{lim}})_{ij} \equiv \lim_{{\bf r}\rightarrow
{\bf r}'}$ $(I_n ({\bf r},{\bf r}',\omega))_{ij}$ ($i,j=x,y,z$).
Note that the (Onsager) reciprocity relation as mentioned in
Section \ref{Bas_eq_Q} implies that $(I_n^{\text{lim}})_{ij} =
(I_n^{\text{lim}})_{ji}$.

\section{The spin flip rate outside a $2$-layer wire}
\label{final formula}

The spin-flip rate in free space is readily derived from
\eq{gamma_B_generall} since $\mbox{Im}[\nabla\times\nabla\times$
$\bm{G}^0({\bf r}_A,{\bf r}_A,\omega_{if} ) ]_{qp}$ $= ( k_3^3/6 \pi
) \, \delta_{qp}$, where $k_3=\omega/c$ is the free space wave
number corresponding to the atomic transition. We use the notation
$k_3$ here because in our discussion of the cylindrical wire, the
third layer is a vacuum. Hence
\bea \label{gamma_0} \Gamma^{\;
0}_{if} =  \, \mu_0  \, \frac{ ( \mu_B g_S )^2 }{3 \pi \, \hbar}
\, k_3^3 S_{if}^{\, 2} \; ,
\eea
where we have introduced the
angular factor $S_{if}^{\, 2} \equiv| \langle i | \hat{S}_x | f
\rangle |^2$ $+ | \langle i | \hat{S}_y | f \rangle |^2 $ $= 2 \,
| \langle i | \hat{S}_x | f \rangle |^2$. We do not have a term
containing $\hat{S}_z$ since we are interested here in a
spin flip, which by definition changes $m_F$. We have moreover
used the fact that the two transverse matrix elements are equal in
absolute value as a result of symmetry. For the $^{87}$Rb ground
state transition $|F,m_F\rangle=|2,2\rangle \rightarrow
|2,1\rangle$, the angular factor $S_{21}^{\, 2} = 1/8$
\footnote{In Ref. \cite{hinds_03}, the angular factor is
erroneously given as $1/10$.}.

In order to find the contribution of the wire to the spontaneous
spin-flip rate, we use  Eqs. (\ref{gamma_B_generall}) and
(\ref{curl_curl_G}). The quantization axis is taken to be along
the $z$ direction, corresponding to the direction of the bias
field that the trapped atoms experience in the experiment.  We
obtain
 \bea \label{gamma_3_dir}
\Gamma^{\text{wire}}_{if}= \frac{3}{8} \, \Gamma^{\; 0}_{if}
\sum_{n=0}^{\infty} ( 2-\delta_{0n} ) \textrm{Re}  \left[
(\widetilde{I}_n^{\text{\text{lim}}})_{xx}
+(\widetilde{I}_n^{\text{lim}})_{yy}   \right]  \,. \eea Here we
have once again used the facts that $|\langle i|\hat{S}_x|f\rangle
|^2=|\langle i|\hat{S}_y|f\rangle |^2$ and that  $\langle i |
\hat{S}_z | f \rangle=0$. The dimensionless integrals
$(\widetilde{I}_n^{\text{lim}})_{ij} \equiv
(I_n^{\text{lim}})_{ij}/k_3^3$ are given by \bea
&&\label{I_xx_Chew} (\widetilde{I}_{n}^{\text{lim}})_{xx} +
(\widetilde{I}_{n}^{\text{lim}})_{yy} = \int_{-\infty}^{\infty}  d
q  \; \frac{1}{\widetilde{\eta}_3^2} \times
\\ \nonumber
&& \hspace*{-3ex} \bigg \{ \,
    [ \, R_n^{11}(q) + q^2 \, R_n^{22}(q) \, ]
    [ \, (H_{n3})^2 \frac{n^2}{k_3^2 (a_2+r)^2}  +
    (\widetilde{\eta}_3 H_{n3}')^2  \, ]
\\ \nonumber
&&  \hspace*{-1ex} + 2 i q  R_{n}^{12}(q)  \left(-\frac{\omega
\varepsilon_3}{k_3} \right) \; \widetilde{\eta}_3  (H_{n3}^2)'
\frac{n}{k_3 (a_2 + r)}  \bigg \} , \eea since $\rho = a_2 + r$.
We have used the simplified notation $Z_{n p} \equiv
Z_n(\widetilde{\eta}_p k_3 \rho)$ and the primes in \eq{I_xx_Chew}
denote the derivative with respect to the full argument of the
relevant function, e.g. $Z_{n p}'\equiv
dZ_n(\widetilde{\eta}_pk_3\rho)/d(\widetilde{\eta}_pk_3\rho)$. We
have also chosen to write the permittivity of the $p$th layer
relative to the outermost layer, i.e. $\varepsilon_p =
\varepsilon_3 \varepsilon_p^{\text{rel}}$. The wave number for
layer $p$ is then given by $k_p^2 = k_3^2
\varepsilon_p^{\text{rel}}$, and the dimensionless propagation
constant $\widetilde{\eta}_p \equiv \eta_p/k_3$ in the $\rho$
direction can be written as $\widetilde{\eta}_p =
\sqrt{\varepsilon_p^{\text{rel}} - q^2}$, where $q \equiv h/k_3$
is the dimensionless integration variable.

The skin depth is the characteristic length scale on which an
electromagnetic wave is damped  within a conducting medium. It is
given by $\delta_p = \sqrt{2 \varepsilon_0 \rho_p \omega}/k_0$
(see e.g. Ref. \cite{Jackson_75}), where $\rho_p$ is the
resistivity of layer $p$. Since the spin-flip frequency is very
much lower than the resonance frequencies  of the material in the
wire, the relative permittivity is related  to the skin depth by
\cite{Jackson_75} \bea  \label{eps_1_APPROX}
\varepsilon_p^{\text{rel}}  \approx  \frac{i}{\varepsilon_0 \rho_p
\omega} =i  \frac{2}{k_0^2  \delta^2_p}  \; . \eea

We see from \eq{G_tot} that the total spin-flip rate is equal to
the sum of the free space contribution and the scattering
contribution. The total spin-flip rate for the rate-limiting
transition $|2,2\rangle \rightarrow |2,1\rangle$ is therefore \bea
\label{gamma_TOT} \Gamma^B_{21} = \, \big ( \Gamma^{\; 0}_{ 21 } +
\Gamma^{\text{wire}}_{ 21 } \, \big ) \, (
\overline{n}_{\text{th}} + 1 )  \; . \eea

\section{Numerical results}
\label{Num_Sec}

In the experiment of Ref.\cite{hinds_03}, cold atoms are held in a
microscopic trap near a current-carrying wire assumed to be at
room temperature. The lifetime for atoms to remain in the
microtrap is measured over a range of distances down to $27 \,
\mu$m from the surface of the wire. The wire consists of a central
copper core with radius $a_1=185 \, \mu$m and a $55 \, \mu$m thick
aluminium layer, i.e. $a_2=240 \, \mu$m. Using \eq{eps_1_APPROX},
the resistivities $\rho_1=1.6\cdot 10^{-8}\,\Omega$m for Cu and
$\rho_2=2.7\cdot 10^{-8}\,\Omega$m for Al give skin depths of
$\delta_1=85 \, \mu$m for Cu and $\delta_2=110\, \mu$m for Al at
frequency $f=\omega/2\pi = 560$ kHz.
\begin{figure}[ht]
\centerline{\includegraphics[width=5.8cm,angle=-90]{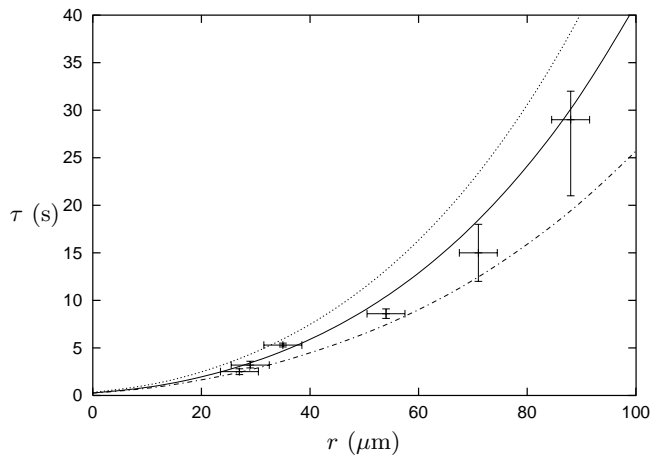}}
\begin{picture}(0,0)
\put(-130,80){$\tau$ (s)}
\put(-10,-6){$r$ ($\mu$m)}
\end{picture}
\caption{ Lifetime $\tau$ of the trapped atom as a function of the
atom-surface distance $r$. {\it Dotted curve}: calculated spin-flip
lifetime near a $2$-layer wire at $300\,$K with the
parameters $f=560\,$kHz, $a_1=185 \, \mu$m, $a_2=240 \, \mu$m,
$\delta_1=85 \, \mu$m, and $\delta_2=110 \, \mu$m. {\it Solid
curve}:  The same but at $380\,$K. {\it Dot-dashed curve}:
calculated lifetime near a thick Al slab at $380\,$K with
$\delta=110 \, \mu$m (using Eq. (35) of Ref. \cite{henkel_99}).
{\it Crosses}: measured lifetimes of Ref. \cite{hinds_03}.}
\label{th_vs_exp_fig}
\end{figure}

The dotted line in \fig{th_vs_exp_fig} shows the lifetime
$\tau=1/\Gamma^B_{21}$ that we have calculated assuming a
temperature of $300\,$K, together with the measured lifetimes
(crosses).  We see that the experimental results are close to the
theory, indicating that the thermal spin flip mechanism is the
primary cause of atom loss in the experiment. Nevertheless there
is also a clear systematic discrepancy, with the measured
lifetimes being $20-30\%$ shorter than expected. We find excellent
agreement when the temperature in our theory is increased to
$380\,$K, as shown by the solid curve in  \fig{th_vs_exp_fig}. We
have re-examined the conditions under which the experiment was run
and consider it most likely that the wire temperature was indeed
$\sim 380\,$K, rather than the $300\,$K previously assumed. Such a
temperature rise would be consistent with known power dissipation
and with reasonable assumptions about the heat flow. In effect,
the thermally driven spin flips have allowed us to measure the
temperature of the wire!

The theory for the decay rate of an atom above a plane, thick slab
is already known \cite{henkel_99}. Applying this theory to an Al
slab with skin depth $\delta=110\,\mu$m and temperature $380\,$K,
we obtain the result shown dot-dashed in \fig{th_vs_exp_fig}. This
curve lies below that for the wire, simply reflecting the fact
that the slab contains a larger volume of fluctuating polarization
than the wire. Naturally, the two $380\,$K curves converge at
sufficiently small atom-surface distances ($r \ll \delta_2$,
$\delta_1$, $a_2$), and in that range they vary linearly with
distance \cite{henkel_99}.

\begin{figure}[ht]
\centerline{\includegraphics[width=5.8cm,angle=-90]{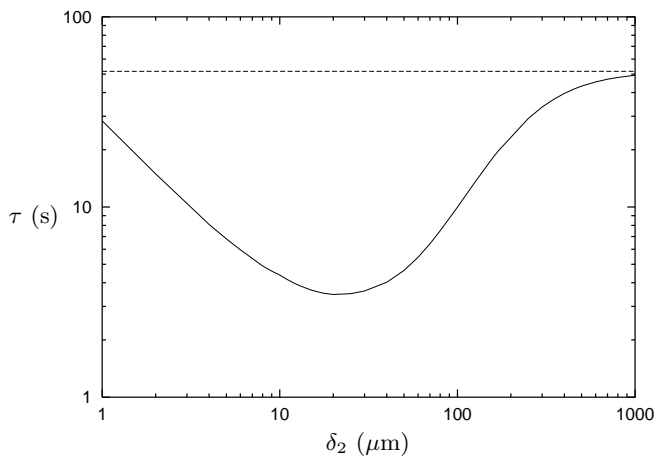}}
\begin{picture}(0,0)
\put(-130,80){$\tau$ (s)} \put(-10,-6){$\delta_2$ ($\mu$m)}
\end{picture}
\caption{ Lifetime $\tau$ of the trapped atom as a function of the
skin depth $\delta_2$ of the  outer layer. The atom-surface
distance is fixed at $r=50 \, \mu$m. The other parameters are:
$f=560$ kHz, $T=300$ K, $a_1=185 \, \mu$m, $a_2=240 \, \mu$m, and
$\delta_1=85 \, \mu$m. The straight dashed line represents the
large $\delta_2$ limit. The numerical value for this limit is
$52\,$s. } \label{delta_2_I_fig}
\end{figure}

The lifetime for the atom to remain in the trap exhibits a minimum
with respect to variation of the skin depth, as illustrated in
Fig.\,\ref{delta_2_I_fig}, where the skin depth of the wire core
$\delta_1$ is fixed at $85\mu$m but the skin depth of the outer
layer $\delta_2$ is varied. Below the minimum at
$\delta_2\simeq20\,\mu$m, a decrease of skin depth leads to an
increase of lifetime in proportion to $\delta_2^{-1}$. This happens
despite a growth in the polarization noise [see Eqs. (\ref{P_N})
and (\ref{eps_1_APPROX})] because the region generating the noise
is becoming thinner. In the small $\delta_2$ limit the outer layer
approaches a perfect conductor, the core wire does not play any
role, and the lifetime becomes exceedingly long. By contrast, when
the skin depth increases above $20\,\mu$m, the reduction in
polarization noise is more influential than the growth of the
source volume.  In this region it is the worse conductor that
gives the longer lifetime. At large $\delta_2$, the outer layer of
the wire approaches the free space limit, and the lifetime is
entirely determined by the skin depth and radius of the core. From
a practical viewpoint it would normally be desirable to avoid the
minimum of the lifetime curve. This means avoiding surface
materials whose skin depth at the spin flip frequency is
comparable with the atom-surface distance. This is a generic
result.
For example, at height z above a slab, one obtains the shortest
lifetime when the skin depth is $z/3^{1/3}$, as is readily derived
from equation (23) of Ref.~\cite{henkel_99}.

\begin{figure}[ht]
\centerline{\includegraphics[width=5.8cm,angle=-90]{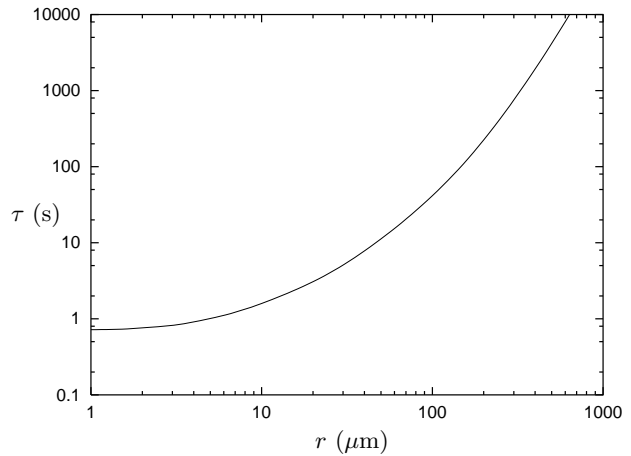}}
\begin{picture}(0,0)
\put(-117,80){$\tau$} \put(-109,80){(s)} \put(-2,-6){$r$ ($\mu$m)}
\end{picture}
\caption{ Lifetime $\tau$ as a function of atom-surface distance
$r$, with $r$, $a_1$ and $a_2$ scaling together according to
$a_2=5r$ and $a_1=(185/240)a_2$. The other parameters are: $f=560$
kHz, $T=300$ K, $\delta_1=85 \, \mu$m, and $\delta_2=110 \, \mu$m.
} \label{miniatur_fig}
\end{figure}

From the same perspective of cold atoms trapped above small
integrated circuits (atom chips) it is also interesting to see how
the lifetime is altered when the dimensions $a_1$, $a_2$ of the
wire are varied or the atom-surface distance $r$ changes. For
example, let us scale all three lengths together, such that
$r=a_2/5$ and $a_1=(185/240)a_2$, while the skin depths are fixed.
The result of such a scaling is illustrated in \fig{miniatur_fig}.
When the atom-surface distance is large compared with the skin
depth, i.e. $r = a_2/5 \gg \delta_2 \sim 100\,\mu$m, the spin-flip
lifetime scales as $\sim r^4$.
This has the same exponent as the $z^4$ scaling of lifetime that
applies at distance $z$ from a slab in the range where $z\gg\delta$
\cite{henkel_99}. The correspondence seems natural to us since the
wire is essentially a curved slab when the skin depth is small.  For a
given ratio of atom-surface distance to wire size, the two lifetimes
should therefore be related by a constant geometrical factor,
resulting in the same distance scaling.
At the opposite extreme, where $z \ll \delta$,
the slab result is $\tau \propto z$. By contrast, we see in
\fig{miniatur_fig} that the lifetime outside the wire approaches a
 constant when $a_2=5r\ll \delta_2$. This difference
occurs because the thickness of the source region is not the skin
depth, but rather the diameter of the wire, which we are scaling
linearly with $r$. In a similar way, it is possible to lengthen
the lifetime of an atom above a slab by reducing the thickness of
the slab to less than the skin depth \cite{hinds_03}.

\begin{figure}[ht]
\centerline{\includegraphics[width=5.8cm,angle=-90]{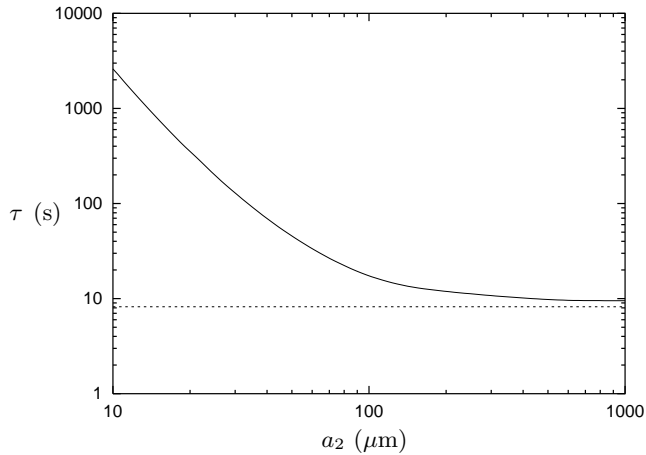}}
\begin{picture}(0,0)
\put(-126,80){$\tau$}
\put(-117,80){(s)}
\put(-8,-6){$a_2$ ($\mu$m)}
\end{picture}
\caption{Lifetime $\tau$ as a function of outer wire radius $a_2$
with the atom-surface distance fixed at $r=50\mu$m. The inner
radius is scaled according to $a_1=(185/240)a_2$. Other parameters
are: $f=560$kHz, $T=300$K, $\delta_1=85\mu$m, and
$\delta_2=110\mu$m. Dotted line: the large $a_2$ limit.}
\label{a_2_fig}
\end{figure}

As a second example of scaling, we change the diameter of the
wire, keeping $a_1=(185/240) a_2$ but fixing the distance from the
surface at $50\,\mu$m. Once again the skin depths are fixed. The
resulting variation in the lifetime of the atom with wire size is
shown in \fig{a_2_fig}. At large wire diameter, the lifetime
approaches $8.2$ s, which is of course the same as the lifetime
$50\,\mu$m above a slab with $110\,\mu$m skin depth. By contrast,
when the wire size is small, i.e. $a_2 \ll r,\delta_2$, the
decreasing volume of material leads to a $a_2^{-3}$ scaling of the
lifetime. In the limit $a_2\rightarrow 0$,
$\Gamma^{\text{wire}}_{if}$ vanishes and the lifetime for the
atoms to remain in the trap is just the free-space blackbody rate
given by the first term in \eq{gamma_TOT}. For $f= 560$ kHz and
$T=300$ K, this free space lifetime is an astonishing $\sim
10^{18}$ s (see also \cite{Purcell}). This figure emphasizes the very
low strength of the electromagnetic field fluctuations in free space
compared with those near a dielectric medium due to the surface modes.

\section{Conclusions}
\label{conclusions}

In this paper we have derived the magnetic spin flip rate for an
atom close to an absorbing dielectric body. The rate is given in
terms of a dyadic Green tensor, allowing the expression to be
applied in principle to a dielectric body of any shape. We derive
an explicit expression for the spin-flip rate of an atom outside a
2-layer cylindrical wire, as used in the experiment of Jones
{\it et al.} \cite{hinds_03}. We compare our numerical results with
their measurements and we find lifetimes marginally longer than
those measured in the experiment. The most likely explanation for
this discrepancy is that the wire was hotter than previously
thought. We also compare the cylindrical case with that of a slab
and show that the spin-flip lifetime is systematically longer
above a cylinder, as one would expect.

We have investigated how the lifetime of the atoms depends on the
skin depth of the material. We find the generic result that there
is a minimum in the lifetime when the skin depth is comparable
with the atom-surface distance. When the dimensions of the wire
and the atom-surface distance $r$ are varied together, the
lifetime scales as $r^4$ at large $r$, following the same scaling
law as a corresponding plane, thick slab, whereas the lifetime
approaches a constant at small $r$. If instead we fix the
atom-surface distance and vary only the dimensions of the wire,
the lifetime scales as $r^{-3}$ when the wire is small, leaving
only the very weak free-space decay rate in the limit of a
vanishing wire diameter. The main conclusion for atom chip design
is that one should avoid a material whose skin depth at the spin
flip transition frequency is comparable with the atom-surface
distance. The lifetime can also be improved by making sure that
metal films on the surface are thinner than the skin depth.

\acknowledgments We are indebted to M.P.A. Jones for valuable
helpful comments. This work was supported by the UK EPSRC and by
the FASTnet and Qgates networks of the EU. P.K.R. acknowledges
support by the Research Council of Norway. S.S. acknowledges
support by the Alexander von Humboldt foundation.

\appendix

\section{The scattering reflection coefficients}
\label{R_appendix}

 The scattering reflection coefficients for a
cylindrical geometry can be computed for any number  of layers
(see e.g. Refs. \cite{chew_90,xiang_96,li_00,tai_93}). In this
Appendix we present the explicit expressions for the scattering
reflection coefficients corresponding to our $3$-layer cylindrical
geometry. To find these reflection coefficients we have used the
iteration tensor equations in Ref. \cite{chew_90}. These iteration
equations are given for arbitrary complex permittivity
$\varepsilon_p$ and arbitrary complex permeability $\mu_p$.
Therefore, the reflection coefficients presented in this Appendix
apply to arbitrary $\varepsilon_p$ {\it and} arbitrary $\mu_p$.
However, we stress that the theory presented in the main body of
this paper is particular to non-magnetic media; we assume that
$\mu_p=\mu_0$ in all the layers $p$.

The reflection coefficients are given as follows:
\bea\label{R_11_final}
R_n^{11}(h)  &=&  \frac{(-1)}{ d_{n  32} } \left[
a^{ H_{3}'  J_2 }_{n  \mu_3}
a^{ J_{3}'  J_2 }_{n  \varepsilon_3}
+ b^{ H_{3} J_{2} }_n b^{ J_{3} J_{2} }_n \right]
\nonumber \\ &+&
 \bigg (  \frac{2  \omega }{\pi  a_2}  \bigg )^2
\eta_3^2  \eta_2^2  \varepsilon_3
\frac{ T_n^{11} }{N_n} \, ,
\\
\label{R_12_final}
R_n^{12}(h)  &=&  \frac{1}{ d_{n  32} } \left[
a^{ H_{3}'  J_2 }_{n \mu_3} b^{ J_{3} J_{2} }_n
- a^{  J_{3}' J_{2} }_{n \mu_3} b^{  H_{3}J_{2}  }_n  \right]
\nonumber \\
&+&  \bigg (  \frac{2  \omega }{\pi  a_2}  \bigg )^2
\eta_3^2  \eta_2^2  \mu_3    \frac{T_n}{N_n} \, ,
\eea
where
\bea
T_n^{11} &=&  d_{n32}\alpha_n -t_{n21} \beta_n \, , \\
T_n \; &=& d_{n32} \gamma_n-t_{n21} \delta_n   \, ,
\eea
and
\bea \nonumber
N_n  &=& (d_{n32} )^2
\bigg[ \,  d_{n  32}  d_{n  21}  +   t_{n  21}  t_{n  32}
\nonumber \\
&-& (a_{11} b_{11}-2\frac{\varepsilon_2}{\mu_2} a_{12} b_{12}+a_{22}
b_{22} ) \, \bigg ] \, .
\eea
Moreover, we have
\bea \label{alpha_a_def}
\alpha_n  &=&-  ( a^{ H_3'  J_2 }_{n  \mu_3} )^2
\varepsilon_2   b_{11}   +  ( b^{  H_3 J_2  }_n )^2\mu_2   b_{22}
\nonumber \\
&& - 2  \varepsilon_2  b_{12} a^{ H_3' J_2 }_{n \mu_3} b^{  H_3 J_2  }_n \, ,
\\  \label{beta_a_def}
\beta_n  &=&- ( a^{ H_3' J_2 }_{n \mu_3} )^2 \varepsilon_2   a_{22}
+ ( b^{ H_3 J_2  }_n )^2  \mu_2  a_{11}
\nonumber \\
&& +2 \varepsilon_2 a_{12} a^{ H_3'J_2 }_{n\mu_3} b^{H_3 J_2}_n  \, ,
\\  \label{gamma_def}
\gamma_n&=& - a^{  H_3'  J_2 }_{n  \mu_3} b^{  H_3'  J_2 }_n
\varepsilon_3  b_{11}
- a^{  H_3'  J_2 }_{n \varepsilon_3} b^{  H_3 J_2  }_n
\mu_2   b_{22}
\nonumber \\
&& + \varepsilon_2  b_{12} [
 a^{ H_3'J_2}_{n\mu_3}  a^{ H_3'  J_2 }_{n\varepsilon_3}
- (  b^{ H_3 J_2}_n )^2  ] \,  ,
\\  \label{delta_def}
  \delta_n &=&-   a^{  H_3'  J_2 }_{n \mu_3} b^{H_3 J_2}_{n\varepsilon} a_{22}
-a^{  H_3'  J_2 }_{n \varepsilon_3} b^{ H_3 J_2  }_n \mu_2  a_{11}
\nonumber \\
&& - \varepsilon_2  a_{12} [ a^{H_3' J_2}_{n\mu_3} a^{H_3'J_2}_{n
\varepsilon_3}  - ( b^{H_3 J_2}_n )^2  ] \,  ,
\eea
and
\bea \label{a_11_def}
a_{11} &=&  a^{ H_3'  J_2 }_{n  \mu_3}
a^{ H_3'  H_2 }_{n  \varepsilon_3}
+b^{ H_3 J_2 }_n b^{ H_3 H_2 }_n \, ,
\\  \label{a_12_def}
a_{12} &=&a^{ H_3'  J_2 }_{n  \mu_3} b^{ H_3 H_2 }_n
-a^{  H_3' H_2 }_{n  \mu_3} b^{  H_3J_2  }_n \nonumber \\
  &=& - \frac{2  \omega}{\pi a_2}  \eta_3^2
\frac{hn}{a_2}   \mu_2  (  H_{n  3}  )^2
\big(k_2^2  - k_3^2\big ) \, ,
\\ \label{b_11_def}
b_{11} &=& a^{ H_2'  J_1 }_{n  \mu_2} a^{ J_2'  J_1 }_{n  \varepsilon_2}
+ b^{ H_2 J_1 }_n b^{ J_2 J_1 }_n   \, ,
\\  \label{b_12_def}
b_{12} &=&a^{ H_2'  J_1 }_{n  \mu_2} b^{ J_2 J_1 }_n
- a^{  J_2' J_1 }_{n  \mu_2} b^{  H_2J_1  }_n
\nonumber \\ &=&-  \frac{2  \omega}{\pi a_1}  \eta_1^2
\frac{hn}{a_1}  \mu_2  (J_{n1})^2 \big(k_1^2  - k_2^2\big ) \, .
\eea
The function $a_{21}$, $a_{22}$, and  $b_{21}$, $b_{22}$ are obtained
from $a_{12}$, $a_{11}$, and $b_{12}$, $b_{11}$, respectively, by
replacing  $\mu_p\leftrightarrow -\varepsilon_p$.
In the last step in Eqs.~(\ref{a_12_def}) and (\ref{b_12_def}) we
have used  the Wronskian determinant between Bessel and Hankel
functions. Finally, we have
\bea
t_{n  21} &=&  a^{ J_2' J_1 }_{n  \mu_2} a^{ J_2' J_1 }_{n  \varepsilon_2}
+ (b^{J_2J_1}_n)^2   , \\
t_{n  32} &=&a^{ H_3' H_2 }_{n  \varepsilon_3} a^{ H_3' H_2 }_{n  \mu_3}
+ (b^{H_3H_2}_n)^2 \, ,
\eea
\bea
d_{n  21} &=&  a^{ H_2'  J_1 }_{n  \mu_2}
a^{ H_2'  J_1 }_{n  \varepsilon_2}
+\big (b^{ H_2  J_1 }_n  \big )^2  \, ,   \\
d_{n  32} &=& a^{ H_3'  J_2 }_{n  \mu_3} a^{ H_3'  J_2 }_{n  \varepsilon_3}
+\big (  b^{  H_3  J_2 }_n  \big )^2   ,
\eea
and
\bea \label{a_HJ_21_mu}
a^{H_2'J_1}_{n  \mu_2} &=& i \omega
\eta_2 \eta_1 \left( \mu_2 \eta_1  H'_{n  2} J_{n  1}
- \mu_1 \eta_2  H_{n  2} J'_{n  1}  \right) ,
\\  \label{a_JJ_21_mu}
a^{J_2'J_1}_{n  \mu_2}  &=&
i \omega \eta_2 \eta_1 \left( \mu_2 \eta_1  J'_{n  2} J_{n  1}
-\mu_1 \eta_2  J_{n  2} J'_{n  1} \right) \, ,
\\ \label{a_HJ_21_eps}
a_{n  \varepsilon_2}^{H_2' J_1} &=& i \omega
 \eta_2 \eta_1  \left(  \varepsilon_2  \eta_1  H'_{n   2} J_{n  1}
 -\varepsilon_1  \eta_2  H_{n  2}  J'_{n  1} \right) \, ,
\\ \label{a_HJ_21}
a_{n  \varepsilon_2}^{J_2' J_1}
&=&  i \omega  \eta_2 \eta_1 \left( \varepsilon_2\eta_1  J'_{n 2} J_{n1}
-\varepsilon_1  \eta_2  J_{n  2} J'_{n  1} \right) \, ,
\\ \label{b_HJ_21}
b_{n}^{H_2 J_1} &=& \frac{hn}{a_1} H_{n2} J_{n1} \left( k_1^2-k_2^2\right).
\eea
Whenever the combination $Z_2$ and $Z_1$ is involved in the
superscript, the radius $a_1$ is implicit in the cylindrical
functions. For example, in Eqs.~(\ref{a_HJ_21_mu})--(\ref{b_HJ_21}) we
have
\bea
&& Z_{n1}\equiv  Z_{n}(\eta_1 a_1) \, ,
Z_{n2}\equiv Z_{n}(\eta_2 a_1)\, ,  \\
&& Z'_{n1} \equiv  \frac{ d Z_{n}(\eta_1 a_1) }{d (\eta_1 a_1)} \, , \quad
Z'_{n2} \equiv \frac{ d Z_{n}(\eta_2 a_1) }{d (\eta_2 a_1)} \, .
\eea
The functions $a^{H_3'J_2}_{n\mu_3}$, $a^{J_3'J_2}_{n\mu_3}$,
$a_{n\varepsilon_3}^{H_3'J_2}$, $a_{n\varepsilon_3}^{H_3'H_2}$,
$a_{n\varepsilon_3}^{J_3'J_2}$, $b_{n}^{H_3J_2}$, and $b_{n}^{H_3H_2}$
are defined analogously, where we understand that the radius $a_2$ is
implicit in all those functions.
Of course, for the special case $\mu_p=1$ for all layers $p$, these
reflection coefficients simplify.

The reflection coefficients $R_n^{21}(h)$ and $R_n^{22}(h)$ can be
obtained from $R_n^{12}(h)$ and $R_n^{11}(h)$, respectively, by
replacing $\mu_p\leftrightarrow-\varepsilon_p$. Note that the
scattering coefficients $R_n^{11}(h)$ as well as 
$R_n^{22}(h)$ are dimensionless. However, the coefficients
$R_n^{12}(h)$ and $R_n^{21}(h)$ are not, but the particular
combinations
$R_n^{12}(h)(-\omega\varepsilon_3/k_3) = R_n^{21}(h)(\omega \mu_3/k_3)$ are.



\begin{thebibliography}{xx}
\bibitem{weinstein_95}
E.A.~Hinds and I.A.~Hughes, J. Phys. D: Appl. Phys. \textbf{32}, R119
(1999);
R.~Folman, P.~Kr\"{u}ger, J.~Schmiedmayer, J.~Denschlag, and C.~Henkel,
Adv. At. Mol. Opt. Phys. {\bf 48}, 263 (2002).
\bibitem{calarco_00}
T.~Calarco, E.A.~Hinds, D.~Jaksch, J.~Schmiedmayer, J.I.~Cirac,
and P.~Zoller, Phys. Rev. A. 61, 022304 (2000).
\bibitem{johnson_28}
J.B.~Johnson, Phys. Rev. {\bf 32}, 97 (1928);
H.~Nyquist, Phys. Rev. {\bf 32}, 110 (1928).
\bibitem{hinds_03}
M.P.A.~Jones, C.J.~Vale, D.~Sahagun, B.V.~Hall, and E.A.~Hinds,
Phys. Rev. Lett. {\bf 91}, 080401 (2003).
\bibitem{harber_03}
D.M.~Harber, J.M.~McGuirk, J.M.~Obrecht, and E.A.~Cornell,
J. Low. Temp. Phys. 133, 229-238  (2003).
\bibitem{henkel_99}
C.~Henkel, S.~P\"{o}tting and M.~Wilkens,
Appl. Phys. B {\bf 69}, 379 (1999);
C.~Henkel and M.~Wilkens,
Europhys. Lett. {\bf 47}, 414 (1999).
\bibitem{dung_00}
L.~Kn\"{o}ll, S.~Scheel, and D.-G.~Welsch, \textit{QED in dispersing
and absorbing media}, in \textit{Coherence and Statistics of Photons
and Atoms}, ed. J. Pe\v{r}ina (Wiley, New York, 2001);
S.~Scheel, L.~Kn\"{o}ll, D.-G.~Welsch, and S.M.~Barnett,
Phys. Rev. A {\bf 60}, 1590 (1999);
S.~Scheel, L.~Kn\"{o}ll  and D.-G.~Welsch,
Phys. Rev. A {\bf 60}, 4094 (1999);
Ho Trung Dung, L.~Kn\"{o}ll  and D.-G.~Welsch,
Phys. Rev. A {\bf 62}, 053804 (2000).
\bibitem{agarwal_75}
G.S.~Agarwal, Phys. Rev. A {\bf 11}, 230 (1975);
J.M.~Wylie and J.E.~Sipe, Phys. Rev. A {\bf 30}, 1185 (1984).
\bibitem{landau_60}
L.D.~Landau and E.M.~Lifshitz, \textit{Electrodynamics of
continuous media} (Pergamon Press, Oxford, 1960).
\bibitem{onsager_31}
L. Onsager,
Phys. Rev. {\bf 37}, 405 (1931); {\bf 38}, 2265 (1931);
\bibitem{Eckhardt}
W.~Eckhardt, Opt. Commun. \textbf{41}, 305 (1982);
Phys. Rev. A \textbf{29}, 1991 (1984).
\bibitem{chew_90}
W.C.~Chew,
\textit{Waves and Fields in Inhomogeneous Media} 
(IEEE Press, New York, 1990).
\bibitem{Jackson_75}
J.D.~Jackson, \textit{Classical Electrodynamics}, 2nd edn. (Wiley,
New York, 1975).
\bibitem{Purcell}
E.M.~Purcell, Phys. Rev. \textbf{69}, 681 (1946).
\bibitem{tai_93}
C.-T.~Tai,
\textit{Dyadic Green Functions in Electromagnetic Theory} 
(IEEE Press, New York, 1993).
\bibitem{xiang_96}
Z.~Xiang and Y.~Lu,
IEEE Trans. Microwave Theory Tech. {\bf 44}, 614 (1996).
\newline
As pointed out in Ref. \cite{li_00}, this paper contains some critical
mistakes.
The relation between the scattering coefficients and the Green tensor
in this paper is inconsistent.
\bibitem{li_00}
L.-W.~Li, M.-S.~Leong, T.-S.~Yeo, and P.-S.~Kooi,
J. of Electromagnetic Waves and Appl. {\bf 14}, 961 (2001).
\end{thebibliography}
\end{document}